\documentclass[twocolumn,superscriptaddress,amssymb,amsmath,nobibnotes,aps,prd,showpacs,
nofootinbib]
{revtex4}%
\usepackage{graphicx}
\usepackage{epsf}
\usepackage{bm}
\usepackage{amsmath}
\usepackage{amsfonts}
\usepackage{amssymb}
\usepackage{epstopdf}
\usepackage{natbib}%
\usepackage{rotating}
\usepackage{pdflscape}
\usepackage{adjustbox}
\pdfoutput=1

\providecommand{\U}[1]{\protect\rule{.1in}{.1in}}
\newcommand{\be}{\begin{equation}}
\newcommand{\ee}{\end{equation}}

\newcommand{\mincir}{\raise
-3.truept\hbox{\rlap{\hbox{$\sim$}}\raise4.truept\hbox{$<$}\ }}
\newcommand{\magcir}{\raise
-3.truept\hbox{\rlap{\hbox{$\sim$}}\raise4.truept\hbox{$>$}\ }}

\usepackage{color}

\begin{document}

\title{Observational Constraints on $f(T)$ gravity from varying  fundamental constants}

\author{Rafael C. Nunes}
\email{nunes@ecm.ub.edu}
\affiliation{Departamento de F\'isica, Universidade Federal de Juiz de Fora, 36036-330,
Juiz de
Fora, MG, Brazil}

\author{Alexander Bonilla}
\email{abonillar@udistrital.edu.com}
\affiliation{Departamento de F\'isica, Universidade Federal de Juiz de Fora, 36036-330,
Juiz de Fora, MG, Brazil}

\author{Supriya Pan}
\email{span@iiserkol.ac.in}
\affiliation{Department of Physical Sciences, Indian Institute of Science Education and
Research -- Kolkata, Mohanpur -- 741246, West Bengal, India}

\author{Emmanuel N. Saridakis}
\email{Emmanuel\_Saridakis@baylor.edu}
\affiliation{Instituto de F\'{\i}sica, Pontificia Universidad de Cat\'olica de
Valpara\'{\i}so,
Casilla 4950, Valpara\'{\i}so, Chile}
\affiliation{Physics Division,
National Technical University of Athens, 15780 Zografou Campus,
Athens, Greece}
\affiliation{CASPER, Physics Department, Baylor University, Waco, TX 76798-7310, USA}


\begin{abstract}
We use observations related to the variation of fundamental constants, in order to impose 
constraints on the viable and most used $f(T)$ gravity models. In particular, for the 
fine-structure constant we use direct measurements obtained by different spectrographic 
methods, while for the effective Newton's constant we use a model-dependent 
reconstruction, using direct observational Hubble parameter data, in order to  
investigate 
its   temporal evolution. We consider two $f(T)$ models and we quantify their 
deviation from $\Lambda$CDM cosmology through a sole parameter. Our analysis reveals that
this parameter can be slightly different from its $\Lambda$CDM 
value, however the best-fit value is very close to the $\Lambda$CDM one. Hence,   
$f(T)$ gravity is consistent with observations, nevertheless, as every modified gravity, 
it may  exhibit only small deviations from $\Lambda$CDM cosmology, a 
feature that must be taken into account in any $f(T)$ model-building.
\end{abstract}
\maketitle

\section{Introduction}
\label{sec:intro}

Modified gravity \cite{Capozziello:2011et} is one of the two main roads one can follow
in order to provide an explanation for the early and late-time universe acceleration
(the second one in the introduction of the dark energy concept
\cite{Copeland:2006wr}). Furthermore, apart from the cosmological motivation,
modified gravity has a theoretical motivation too, namely to improve the
renormalizability properties of standard general relativity \cite{Stelle:1976gc}.

In constructing a gravitational modification, one usually starts from the
Einstein-Hilbert action and extends it accordingly. Thus, he can obtain $f(R)$ gravity
\cite{Nojiri:2010wj}, Gauss-Bonnet and $f(G)$ gravity
\cite{Nojiri:2005jg},   gravity with higher-order curvature invariants
\cite{Naruko:2015zze},  massive gravity \cite{deRham:2014zqa} etc. Nevertheless, he could
start from the equivalent, torsional formulation of gravity, namely from the
Teleparallel Equivalent of General Relativity (TEGR)
\cite{ein28,Hayashi79,Pereira.book}, in which the
gravitational Lagrangian is the torsion scalar $T$, and construct various modifications,
such as $f(T)$ gravity
\cite{Bengochea:2008gz,Linder:2010py,Dent:2011zz,Geng:2011aj,Bamba:2013jqa} (see
\cite{Cai:2015emx} for a review), teleparallel Gauss-Bonnet gravity
\cite{Kofinas:2014owa}, gravity with higher-order torsion invariants
\cite{Otalora:2016dxe}, etc.

An important question in the above gravitational modifications is what are the forms of
the involved unknown functions, and what are the allowed values of the various
parameters. Excluding forms and parameter regimes that lead to obvious
contradictions and problems, the main tool we have in order to provide further
constraints is to use observational data. For the case of torsional gravity one can use
solar system data \cite{Iorio:2012cm}, or cosmological
observations from Supernovae type Ia, cosmic microwave background and baryonic acoustic
oscillations \cite{Wu:2010mn,Nesseris:2013jea,Nunes:2016qyp}.

On the other hand, in some modified cosmological scenarios one can obtain a variation
of the fundamental constants, such as the fine structure and the Newton's constants. Such
a possibility has been investigated in the literature since Dirac \cite{dirac} and Milne
and Jordan \cite{milne-jordan} times. Later on, Brans and Dicke proposed the time
variation of the  Newton's constant, driven by a dynamical scalar field
coupled to curvature \cite{brans-dicke}, while Gamow  triggered subsequent speculations
on the possible variation of the fine structure constant \cite{Gamow}. Similarly, in
recent modified gravities, which involve extra degrees of freedom comparing to general
relativity, one may obtain such a variation of the fundamental constants
\cite{Dvali:2001dd,Bento:2004jg}.
However, since experiments and observations give strict bounds on these variations
\cite{Magueijo:2003gj,Avelino:2006gc,Sola:2015xga,Stadnik:2014tta}, one
can use them in order to constrain the theories at hand.

In the present work we are interested in investigating the constraints on $f(T)$ gravity
by observations related to the variation of fundamental constants. In particular, since
$f(T)$ gravity predicts a variation of the fine-structure and Newton's constants, we will
use the recent observational bounds of these variations in order to constrain the $f(T)$
forms as well as the range of the involved parameters. The plan of the work is the
following: In Section \ref{fTcosmology} we give a brief review of $f(T)$ gravity and
cosmology. In Section \ref{alpha} we investigate the constraints on specific $f(T)$
gravity models arising from  the observational bounds of the fine-structure constant
variation,  while in Section \ref{G} we study the corresponding constraints that arise
from the observational bounds of the Newton's constant variation. Finally, in Section
\ref{Conclusions} we summarize our results.

\section{ $f(T)$ gravity and cosmology}
\label{fTcosmology}

In this section we provide a short review of $f(T)$ gravity and cosmology. We use the
tetrad fields $e^\mu_A$, which form an orthonormal base at each point of the
tangent space of the underlying manifold $(\mathcal{M},g_{\mu\nu})$,
where $g_{\mu\nu}=\eta_{A B} e^A_\mu e^B_\nu$ is the metric tensor defined on
this manifold (we use Greek indices for the coordinate space and
Latin indices for the tangent one). Furthermore,
instead of the torsionless Levi-Civita connection
which is used in the Einstein-Hilbert action, we use the curvatureless
Weitzenb{\"{o}}ck connection $\overset{\mathbf{w}}{\Gamma}^\lambda_{\nu\mu}\equiv
e^\lambda_A\:
\partial_\mu e^A_\nu$  \cite{Pereira.book}. Hence, the gravitational field
in such a formalism is described by the following torsion tensor:
\begin{equation}
T^\rho_{\verb| |\mu\nu} \equiv e^\rho_A
\left( \partial_\mu e^A_\nu - \partial_\nu e^A_\mu \right).
\end{equation}
Subsequently, the Lagrangian of the Teleparallel equivalent of general relativity, namely 
the
torsion scalar $T$, is constructed by contractions of the torsion tensor as
\cite{Pereira.book}
\begin{equation}
\label{Tscalar}
T\equiv\frac{1}{4}
T^{\rho \mu \nu}
T_{\rho \mu \nu}
+\frac{1}{2}T^{\rho \mu \nu }T_{\nu \mu\rho}
-T_{\rho \mu}{}^{\rho }T^{\nu\mu}{}_{\nu}\, \, .
\end{equation}

One may consider generalized theories in which the Lagrangian $T$
is extended to an arbitrary function $f(T)$,  similarly to the $f(R)$ extension of
curvature-based gravity. In particular, such gravitational action will read as
\begin{eqnarray}
\label{actionbasic}
 {\mathcal S}_{gr} = \frac{1}{16\pi G_N}\int d^4x |e| f(T),
\end{eqnarray}
where $e = \text{det}(e_{\mu}^A) = \sqrt{-g}$, and  $G_N$ is the
Newton's constant. Additionally, along the gravitational action (\ref{actionbasic}) we
consider the matter sector, and hence the total action writes as
\begin{align}\label{f(T)-action}
\mathcal{S} & = \frac{1}{16\pi G_N}  \int d^4 x\, |e|\, f(T) + \int d^4 x\,
\mathcal{L}_m
(e^A_{\mu}, \Psi_M),
\end{align}
where $\mathcal{L}_m (e^A_{\mu}, \Psi_M)$ is the total matter Lagrangian including
 the electromagnetic field. Finally, variation in terms of the tetrad
fields give rise to the field equations as
\begin{eqnarray}
\label{field-equations}
&& \!\!\!\!\!\!\!\!\!\!\!\!\!\!
e^{-1}\partial_{\mu}(ee_A^{\rho}S_{\rho}{}^{\mu\nu})f_{T}
 -f_{T}e_{A}^{\lambda}T^{\rho}{}_{\mu\lambda}S_{\rho}{}^{\nu\mu}+\frac{1}{4} e_ { A
} ^ {
\nu
}f({T})
\nonumber \\
&&
\ \ \ \ \ \ \ \ \
+
e_A^{\rho}S_{\rho}{}^{\mu\nu}\partial_{\mu}({T})f_{TT}
= 4\pi G_Ne_{A}^{\rho}
{\mathcal{T}^{(m)}}_{\rho}{}^{\nu},
\end{eqnarray}
where $f_{T}=\partial f/\partial T$, $f_{TT}=\partial^{2} f/\partial T^{2}$,
and with ${\mathcal{T}^{(m)}}_{\rho}{}^{\nu}$ the total matter
energy-momentum tensor. In the above equation  we have inserted for convenience the
``super-potential'' tensor $S_\rho^{\verb| |\mu\nu} = \frac{1}{2}
\left(K^{\mu\nu}_{\verb|  |\rho}+\delta^\mu_\rho
  T^{\alpha \nu}_{\verb|  |\alpha}-\delta^\nu_\rho
 T^{\alpha \mu}_{\verb|  |\alpha}\right)$ defined in terms of the co-torsion tensor
$K^{\mu\nu}_{\verb| |\rho}=-\frac{1}{2}\left(T^{\mu\nu}_{\verb|  |\rho} - T^{\nu
\mu}_{\verb|  |\rho} - T_\rho^{\verb| |\mu\nu}\right)$.

Applying $f(T)$ gravity in a cosmological framework we   consider a
spatially flat FLRW universe with line element $ds^2= -dt^2 + a^2 (t) [dr^2 + r^2\, d
\theta^2 + \sin^2 \theta \, d \phi^2]$, which arises from the diagonal tetrad
$e_{\mu}^A={\rm diag}(1,a(t),a(t),a(t))$, with $a (t)$ the scale factor. In this case,
the field equations (\ref{field-equations}) become
\begin{align}
H^2 & = \frac{8 \pi G_N}{3} \Bigl(\rho+ \rho_T \Bigr),\label{F1.1}\\
\dot{H} & = - 4 \pi G_N\, \Bigl[ \left(p+ p_T  \right)+ \left(\rho+ \rho_T \right)
\Bigr],\label{F2.
1}
\end{align}
where $\rho$ and $p$ are respectively the total matter energy density and pressure, and
where $\rho_T$, $p_T$ are the effective dark-energy energy density and the pressure of 
gravitational
origin, given by
\begin{align}
\rho_T & = \frac{1}{16 \pi G_N}\, \left[ 2 T f_T - f (T) -
T\right],\label{F1.1a}\\
p_T & =  \frac{1}{16 \pi G_N}\, \left[ 4 \dot{H}\, \left( 2T f_{TT} +
f_T-
1\right) \right] - \rho_T~.
\label{F2.1b}
\end{align}
In the above expressions we have used that
\begin{eqnarray}
\label{TH62}
T = - 6H^2,
\end{eqnarray}
 which arises straightforwardly
from (\ref{Tscalar}) in the FLRW universe. Finally, from 
equations (\ref{F1.1a}), (\ref{F2.1b}) we can define 
the effective dark-energy equation of 
state (EoS) as
\begin{align}\label{total-eos}
w = -1 - \frac{2 \dot{H}}{3 H^2}= -1 + \frac{2}{3}H(1+z) \frac{dH}{dz},
\end{align}
where as usual we use the redshift
$z=\frac{a_0}{a}-1$, as the independent variable, and for simplicity we set $a_0=1$. 
Clearly, $w$ has a dynamical nature. 

In the following we focus on two  well-studied, viable $f(T)$ models, which correspond to
a small deviation from $\Lambda$CDM cosmology,  and which according to 
\cite{Wu:2010mn,Nesseris:2013jea,Nunes:2016qyp} are the ones that fit the observational
data very efficiently.

\begin{itemize}

\item

The first scenario is the power-law model  (hereafter $f_{1}$CDM)  introduced in
\cite{Bengochea:2008gz}, with
\begin{eqnarray}
\label{model1}
f (T)   = T + \theta   \left(-T \right)^{b},
\end{eqnarray}
where $\theta$, $b$ are the two free model parameters, out of which only one is
independent. Inserting this  $f(T)$ form into the first Friedmann equation (\ref{F1.1})
at present time, i.e. at redshift $z= 0$,
one may derive that
\begin{eqnarray}
\label{theta}
\theta=(6H_0^2)^{1-b}\left(\frac{1- \Omega_{m0}}{2b-1}  \right),
\end{eqnarray}
where $\Omega_{m0}=\frac{8\pi G \rho_{m0}}{3H_0^2}$ is the corresponding
density parameter at present. Hence, and using additionally that
$\rho_{m}=\rho_{m0}(1+z)^{3}$, equation (\ref{F1.1}) for
this model can be written as
\begin{equation}
\ \ \ \ \ \ \ \ \
\frac{H^2(z)}{H_0^2}  = \left(1- \Omega_{m0}\right)
\left[\frac{H^2(z)}{H_0^2}\right]^b  
+ \Omega_{m0}(1+z)^3 .
\label{E2model1}
\end{equation}
Lastly, we mention that the above model for $b= 0$ reduces to
  $\Lambda$CDM cosmology, while for  $b= 1/2$ it gives rise to the
Dvali-Gabadadze-Porrati (DGP) model \cite{Dvali:2000hr}.

\item

The second scenario is the square-root-exponential (hereafter $f_{2}$CDM) of
 \cite{Linder:2010py}, with
\begin{eqnarray}
\label{model2}
f(T)=T + \beta T_{0}(1-e^{-p\sqrt{T/T_{0}}}),
\end{eqnarray}
in which $\beta$ and $p$  the two free model parameters out of which only one is
independent.
Inserting this  $f(T)$ form into (\ref{F1.1}) at present time, one
obtains that
\begin{eqnarray}
\label{beta}
\beta=\frac{1- \Omega_{m0}}{1-(1+p)e^{-p}}.
\end{eqnarray}
Finally, the first Friedmann equation (\ref{F1.1})  for
this model can be written as
\begin{eqnarray}
&& \ \ \
\frac{H^2(z)}{H_0^2}  + \frac{1-\Omega_{m0} }{1-(1+p)e^{-p}}\!\left\{
\left[1+ \frac{pH(z)}{H_0}    \right]
\, e^{-\frac{pH(z)}{H_0}  }\!- \!1\right\}
\nonumber\\
&& \ \ \ \ \ \ \ \ \ \ \ \ \ \ \ \,
= \Omega_{m0}\,(1+z)^3 .
\label{E2model2}
\end{eqnarray}
Lastly, note that this model reduces to   $\Lambda$CDM cosmology for
$p\rightarrow+\infty$. Hence, in the following, for this model it will be convenient
to set $b\equiv 1/p$, and hence $\Lambda$CDM cosmology is obtained for $b \rightarrow
0^{+}$.

 \end{itemize}

\section{Observational constraints from fine-structure constant variation}
\label{alpha}

In this section we will use observational data of the variation of the fine-structure
constant $\alpha$, in order to constrain $f(T)$ gravity. Let us first quantify the
$\alpha$-variation in the framework of $f(T)$ cosmology. In general, in a
given theory the fine-structure  constant is obtained using the coefficient of the
electromagnetic Lagrangian. In the case of modified gravities, this coefficient generally
depends on the new degrees of freedom of the theory \cite{Olive:2001vz,Bento:2004jg}. 
Even if one starts from the Jordan-frame formulation of a theory, with an uncoupled 
electromagnetic Lagrangian, and although the electromagnetic Lagrangian is conformally 
invariant, and it is not affected by conformal transformations between the Jordan 
and 
Einstein frames, thus it will acquire a dependence on the extra degree(s) of freedom due 
to 
quantum effects \cite{Brax:2012gr}. In particular, if $\phi$ is the extra degree of 
freedom that arises from the conformal transformation  $\tilde{g}_{\mu \nu}  =\Omega^2 
g_{\mu \nu}$ from the Jordan to the Einstein frame, then quantum effects such as the 
presence of heavy fermions (note that this does not necessarily require new physics  
till the Planck scale) will 
induce 
a coupling of $\phi$ to photons, namely 
\cite{Brax:2012gr}
\begin{eqnarray}
 S_{EM}=-\frac{1}{g_{bare}^2}\int d^4x\sqrt{-g}B_F(\phi)F_{\mu\nu}F^{\mu\nu},
\label{Sgauge}
\end{eqnarray}
where $F_{\mu\nu}$ is the electromagnetic tensor, $g_{bare}$ the bare coupling 
constant, and
\begin{eqnarray}
 B_F(\phi)=1+\beta_\gamma \frac{\phi}{M_{pl}}+\cdots,
\label{BF}
\end{eqnarray}
with $M_{pl}=1/(8\pi G_N)$ the Planck mass and $\beta_\gamma={\cal{O}}(1)$ a constant (we 
have assumed that $\beta_\gamma\phi\ll M_{pl}$). Hence, the scalar coupling to the 
electromagnetic field will imply a dependence of the fine structure constant of the form 
\cite{Olive:2001vz,Brax:2012gr}
\begin{eqnarray}
\frac{1}{\alpha_{E}}=\frac{1}{\alpha_{J}} B_F(\phi),
\label{finestuct1}
\end{eqnarray}
where the subscripts denote the Einstein and Jordan frames respectively,
or equivalently
\begin{eqnarray}
\frac{\Delta\alpha}{\alpha}\equiv \frac{\alpha_{E}-\alpha_{J}}{\alpha_{J}}    =\frac{1}{ 
B_F(\phi)} -1.
\label{finestuct1b}
\end{eqnarray}
The above factor is in general time- (i.e redshift-) dependent. Therefore, it proves 
convenient to normalize it in order to have $\Delta\alpha=0$ at present ($z=0$), which in 
case where $B_F(z=0)\equiv B_{F0}\neq1$ is obtained through a rescaling 
$F_{\mu\nu}\rightarrow\sqrt{ B_{F0}} F_{\mu\nu}$ and $B_F\rightarrow B_F/B_{F0}$. Thus, 
we result to 
\begin{eqnarray}
\frac{\Delta\alpha}{\alpha}=\frac{B_{F0}}{
B_F(\phi)} -1.
\label{finestuct1c}
\end{eqnarray}

Although the above procedure is straightforward in cases where a conformal transformation
from the Jordan to the Einstein frame exists, it becomes  more complicated for
theories where such a transformation is not known. In case of $f(T)$ gravity, it is well
known that a conformal transformation does not exist in general, since transforming the
metric as $\tilde{g}_{\mu \nu}  =\Omega^2 g_{\mu \nu}$, with $\Omega^2=f_T$ a smooth
non-vanishing function of spacetime coordinates, one obtains Einstein gravity plus a
scalar field Lagrangian, plus the transformed matter Lagrangian, plus a non vanishing
term   $2\Omega^{-6}\tilde{\partial}^\mu \Omega^2\tilde{T}^\rho_{\ \rho\mu}$
\cite{Yang:2010ji}. This additional term forbids the complete transformation
to the Einstein frame, and hence in every application one has indeed to perform
calculations in the more complicated Jordan one.

In order to avoid   performing calculations in the Jordan frame we will make the 
reasonable
assumption that $f(T)=T+const.+\text{corrections}$, which has been shown to be the case
according to observations \cite{Iorio:2012cm,Wu:2010mn,Nesseris:2013jea,Nunes:2016qyp},
and holds for the two forms considered in this work, namely (\ref{model1}) and
(\ref{model2}), too. Hence, $\Omega^2=1+\text{corrections}$, and then
$\tilde{\partial}^\mu \Omega^2$ is negligible, which implies that the above extra term can
be neglected. In the end of our investigation, we will verify the validity of the above
assumption. Thus, we can indeed obtain an approximate transformation to the Einstein
frame, and in particular the introduced degree of freedom reads as 
$\phi=-\sqrt{3}/f_T$  \cite{Yang:2010ji}. Hence, inserting this into (\ref{BF}) we acquire
\begin{eqnarray}
 B_F(\phi)=1-\frac{\sqrt{3}\beta_\gamma}{M_{pl} f_T}+\cdots,
\label{BF2}
\end{eqnarray}
and thus inserting into relation (\ref{finestuct1c}), we can  
easily extract the  variation of the fine-structure
constant as   
\begin{equation}
\label{fc0}
\frac{\Delta \alpha}{\alpha} (z)= \frac{M_{pl}f_{T0}-\sqrt{3}\beta_\gamma 
}{M_{pl}f_T(z)-\sqrt{3}\beta_\gamma } - 1,
\end{equation}
where $f_{T0}= f_T (z= 0)$. Lastly, since $\beta_\gamma={\cal{O}}(1)$, the above relation 
becomes 
\begin{equation}
\label{fc}
\frac{\Delta \alpha}{\alpha} (z)\approx\frac{f_{T0}}{f_T(z)} - 1.
\end{equation}
Hence, for a general $f(T)$, the ratio $\Delta \alpha /
\alpha$ indeed depends on $z$, through the $f_T(z)$ function (we remind that according to
(\ref{TH62}), $T(z)=-6H^2(z)$), while in the case of standard $\Lambda$CDM cosmology,
where $f(T)=T+\Lambda$, $\Delta \alpha / \alpha$ becomes  zero.
\begin{table}[!h]
      \begin{center}
          \begin{tabular}{cccc}
          \hline
          \hline
               &$ z $&$ \Delta \alpha/ \alpha (ppm) $& Ref.\\
          \hline
          \hline
               &$ 1.08 $&$ 4.3 \pm 3.4$& \cite{Songaila:2014fza}\\
               &$ 1.14 $&$ -7.5 \pm 5.5$& \cite{Evans:2014yva}\\
               &$ 1.15 $&$ -0.1 \pm 1.8$& \cite{Molaro:2007kp}\\
               &$ 1.15 $&$ 0.5 \pm 2.4$& \cite{Chand:2006va}\\
               &$ 1.34 $&$ -0.7 \pm 6.6$& \cite{Evans:2014yva}\\
               &$ 1.58 $&$ -1.5 \pm 2.6$& \cite{Molaro:2013saa}\\
               &$ 1.66 $&$ -4.7 \pm 5.3$& \cite{Songaila:2014fza}\\
               &$ 1.69 $&$ 1.3 \pm 2.6$& \cite{Agafonova:2011sp}\\
               &$ 1.80 $&$ -6.4 \pm 7.2$& \cite{Songaila:2014fza}\\
               &$ 1.74 $&$ -7.9 \pm 6.2$& \cite{Evans:2014yva}\\
               &$ 1.84 $&$ 5.7 \pm 2.7$& \cite{Molaro:2007kp}\\
          \hline
          \hline
          \end{tabular}
      \end{center}
      \caption{Compilation of recent measurements of the fine-structure constant
obtained by different spectrographic methods. For details in each case, see
the corresponding references.}
      \label{tab_data}
\end{table}

In the following, we confront relation (\ref{fc}) with observations of the
fine-structure constant variation, in order to impose constraints on $f(T)$ gravity (it 
proves that the neglected term between (\ref{fc0}) and (\ref{fc}) imposes an error of the 
order of $10^{-9}$ and hence our approximation is justified). We
use direct measurements of the fine-structure constant that are obtained by different
spectrographic methods, summarized in Table \ref{tab_data}. Additionally, along with these
data sets, and in order to diminish the degeneracy between the free parameter of the
models, we use $580$ Supernovae data (SNIa) from Union 2.1
compilation \cite{Suzuki:2011hu}, as well as data from BAO observations, adopting the
three measurements of $A(z)$ obtained in \cite{Blake:2011en}, and using the
covariance among these data given in \cite{Shi:2012ma}.

In the following two subsections, we  analyze two viable models, namely
$f_1$CDM of (\ref{model1}) and $f_2$CDM of (\ref{model2}), separately.

\subsection{Model $f_1$CDM: $f(T)  = T + \theta  \left(-T \right)^{b}$}

For the power-law  $f_1$CDM  model of (\ref{model1}), we easily acquire
\begin{align}
\label{fcdm1}
f_T(z) & = 1- b\,\left(\frac{1-\Omega_{m0}}{2b-1} \right)\,
\left[\frac{H^2(z)}{H_0^2}\right]^{(b-1)},
\end{align}
where we have used also (\ref{TH62}).
Inserting  (\ref{fcdm1})  into (\ref{fc}) we can derive the evolution of
$\Delta \alpha / \alpha$ as
\begin{align}
	\label{fcdm1a1aa}
	\frac{\Delta \alpha}{\alpha} (z) \approx \frac{\left[1-
		b\,\left(\frac{1-\Omega_{m0}}{2b-1} \right)\right]}{\left\{1- 
b\,\left(\frac{1-\Omega_{m0}}{2b-1} 
		\right)\,
		\left[\frac{H^2(z)}{H_0^2}\right]^{(b-1)}\right\}}-1,
	\end{align}
where the ratio $H^2(z)/H_0^2$ is given by (\ref{E2model1}).

We mention that while analyzing the model for the data set of $\Delta
\alpha/ \alpha$ of Table \ref{tab_data}, we have marginalized over $\Omega_{m0}$,
and thus the statistical information focuses only on the parameter $b$. For the fittings
$\Delta \alpha/ \alpha$
 $+$ SNIa and $\Delta \alpha/ \alpha$  $+$ SNIa $+$ BAO, we have considered $\Omega_{m0}$ 
as a free parameter, and we have found that
$\Omega_{m0} = 0.23 \pm 0.13$ (for $\Delta \alpha/ \alpha$ $+$ SNIa) and
$\Omega_{m0} = 0.293 \pm 0.023$ (for $\Delta \alpha/ \alpha$ $+$ SNIa $+$ BAO) at
1$\sigma$ confidence level.
\begin{figure}[ht]
 	\includegraphics[width=3.0in]{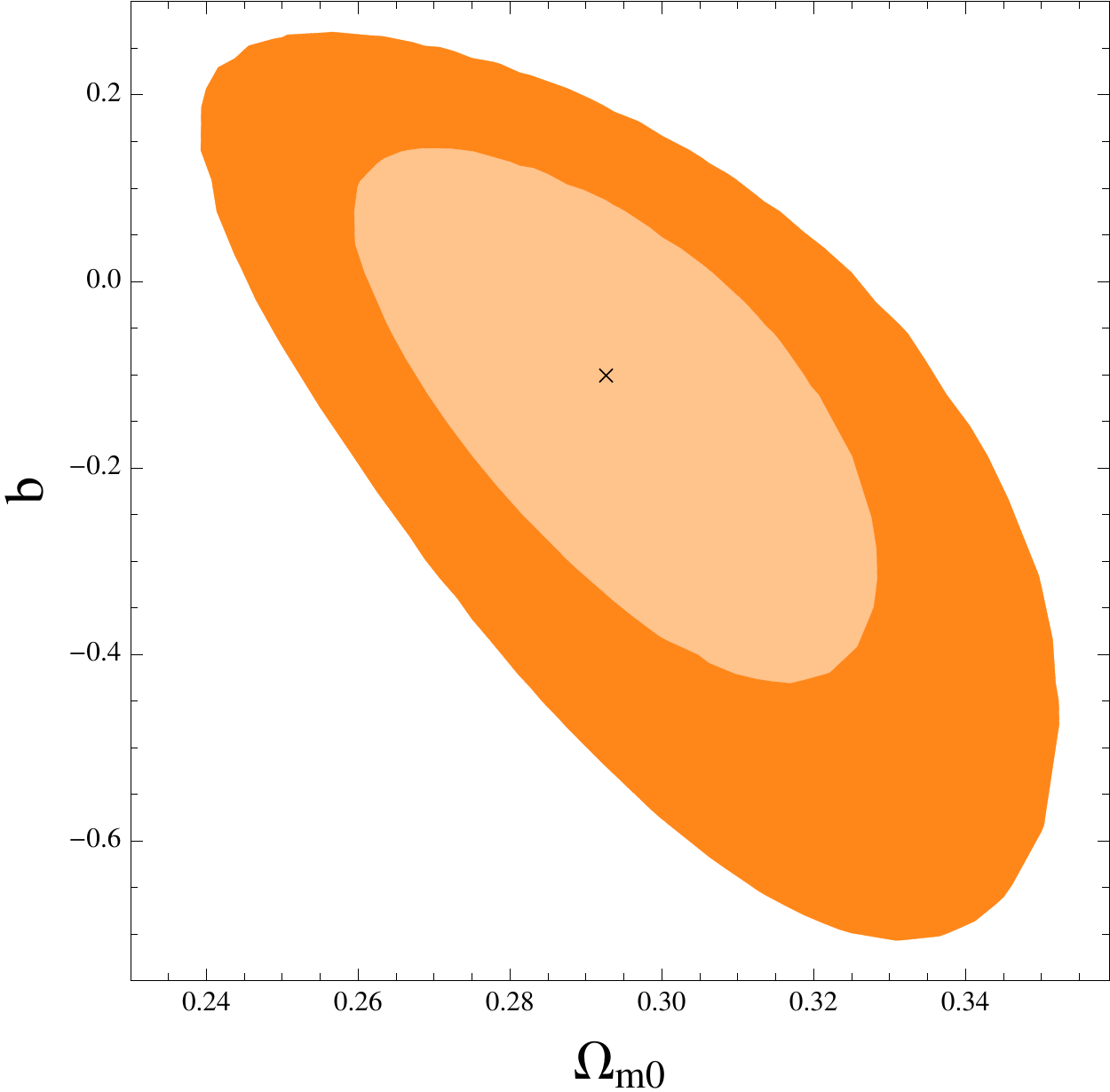}
 	\caption{\label{f1_Delta_alpha_2}  \textit{1$\sigma$ and 2$\sigma$ confidence
regions for the $f_1$CDM  power-law model of (\ref{model1}),  obtained from the joint
analysis   $\Delta \alpha/ \alpha$ $+$ SNIa $+$ BAO. The cross marks the best-fit value.
}}
\end{figure}
Finally, in Fig. \ref{f1_Delta_alpha_2} we present the
68.27$\%$ and 95.45$\%$ confidence regions in the plane $\Omega_{m0} - b$,
considering the observational data $\Delta \alpha/ \alpha$  $+$ SNIa $+$ BAO. Note that
these results are in qualitative agreement with those of different observational fittings
\cite{Wu:2010mn,Nesseris:2013jea}, and show that $\Lambda$CDM cosmology (which is
obtained for $b=0$) is inside the obtained region. In fact, one may notice from 
Table \ref{f1_1} that the reduced $\chi^2$ for $ \Delta \alpha/ \alpha$ $+$ SNIa and 
$\Delta \alpha/ \alpha$ $+$ SNIa $+$ BAO data are very close  to $1$, while for 
single data from $\Delta \alpha/ \alpha$  its value slightly exceeds $1$ although not  
significantly. 
\begin{table}[!h]
      \begin{center}
          \begin{tabular}{cccc}
          \hline
          \hline
               &$ Data $&$ b $& $\chi^2_{min}/d.o.f$\\
          \hline
          \hline
               &$ \Delta \alpha/ \alpha $                   & $0.35$ $\pm$ $0.40$  & 
$1.1$   \\
               &$ \Delta \alpha/ \alpha $+$ SNIa$           & $0.25$ $\pm$ $0.70$   & 
$0.96$   \\
               &$ \Delta \alpha/ \alpha $+$ SNIa $+$ BAO$   & $-0.10$ $\pm$ $0.18$  & 
$0.97$  \\
          \hline
          \hline
          \end{tabular}
      \end{center}
      \caption{ Summary of the best fit values of the parameter $b$  of the
 $f_1$CDM  power-law model   of (\ref{model1}), for three
different observational data  sets with reduced $\chi^2$: $\chi^2_{min}/d.o.f$ ($d.o.f$
stands for the ``degrees of freedom'').}
      \label{f1_1}
\end{table}

\begin{figure}[ht]
 	\includegraphics[width=3.0in]{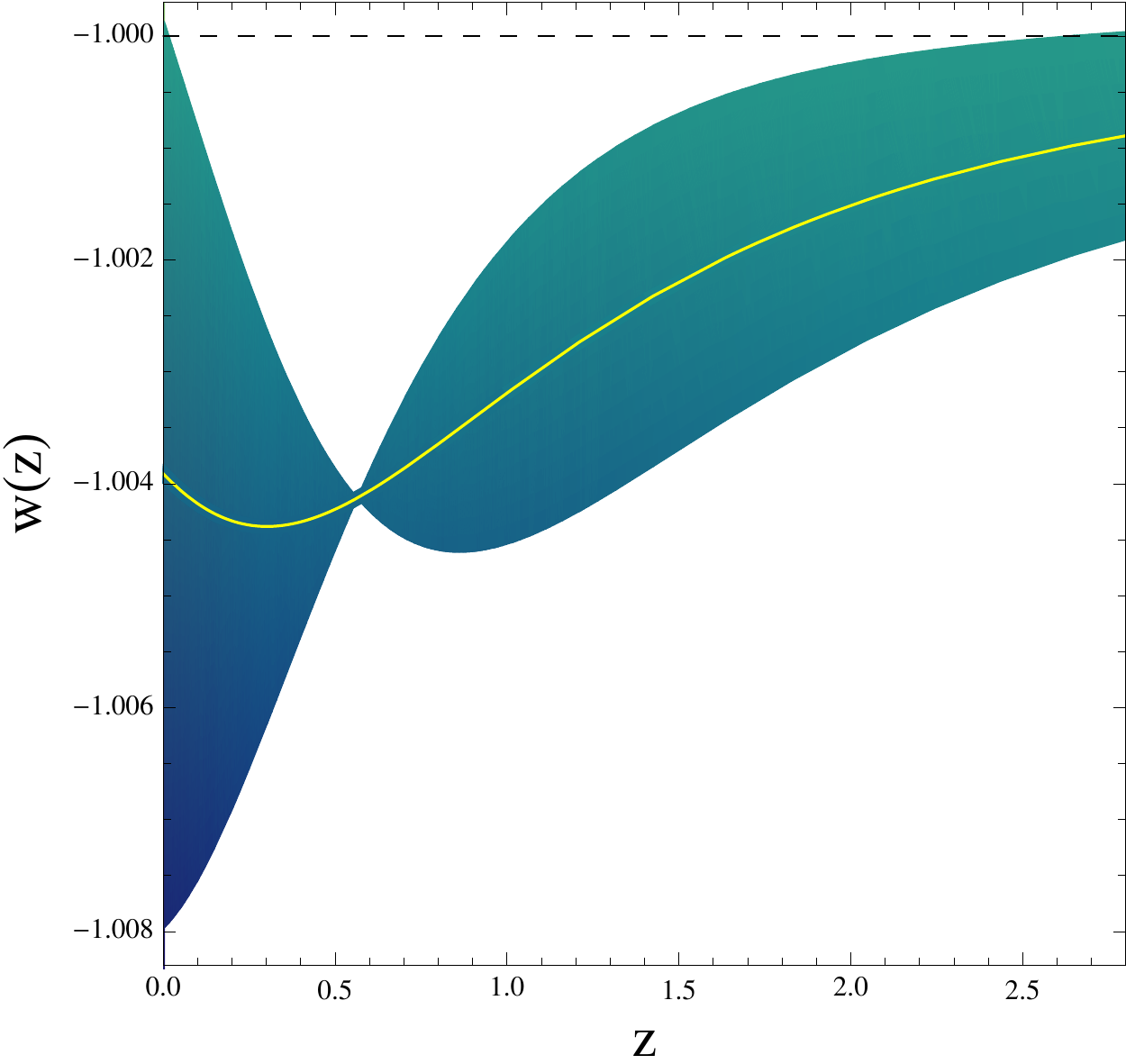}
 	\caption{\label{f1-totalEoS}  \textit{The evolution of the  
equation-of-state parameter given in (\ref{total-eos}),  for the
 $f_1$CDM  power-law model of (\ref{model1}), applying a reconstruction 
at 1$\sigma$ confidence level via error propagation using the joint analysis $\Delta 
\alpha/ \alpha$  $+$ SNIa $+$ BAO.}}
\end{figure}

Additionally, in order to examine the late-time asymptotic behavior of the 
scenario at hand, in Fig. \ref{f1-totalEoS} we depict the evolution of the  
equation-of-state parameter given in (\ref{total-eos}), applying a reconstruction 
at 1$\sigma$ confidence level via error propagation using the joint analysis $\Delta 
\alpha/ \alpha$  $+$ SNIa $+$ BAO. As we can see,  $w$ at late times acquires values very 
close to `$-1$', as expected. For a more detailed investigation of the  late-time 
asymptotics up to the far future one must apply the method of dynamical system analysis 
as it was done in \cite{Wu:2010xk}, where it was thoroughly shown that the universe 
will end in a de Sitter phase.  
 
In summary, it is clear that $f_1$CDM model, under all the 
above three different combinations of statistical data sets, remains close to 
$\Lambda$CDM cosmology as expected. Lastly, note that this is a self-consistent 
verification for the validity of our assumption that  $f(T)=T+const.+\text{corrections}$, 
which allowed us to work in the Einstein frame.

\subsection{Model $f_2$CDM: $f(T)=T + \beta T_{0}(1-e^{-p\sqrt{T/T_{0}}})$}

For the square-root-exponential  $f_2$CDM  model of (\ref{model2}), we easily obtain
\begin{eqnarray}\label{fcdm2}
f_T (z) = 1+ \frac{p}{2} \left[ \frac{1-\Omega_{m0}}{1-(1+p)\,
e^{-p}}\right]\,
\left[ \frac{
H_0}{H(z)} \right] e^{-\frac{pH(z)}{H_0}}.
\end{eqnarray}
Inserting  (\ref{fcdm2})  into (\ref{fc}) we can derive the evolution of
$\Delta \alpha / \alpha$ as

	\begin{align}
	\label{fcdm1a2bb}
	\frac{\Delta \alpha}{\alpha} (z) \approx
	\frac{\left\{
		1+ \frac{p}{2}\left[ \frac{1-\Omega_{m0}}{1-(1+p)
			e^{-p}}\right] e^{-p}
		\right\}}{\left\{ 1+ \frac{p}{2} \left[ \frac{1-\Omega_{m0}}{1-(1+p)\,
			e^{-p}}\right]\,
		\left[ \frac{
			H_0}{H(z)} \right] e^{-\frac{pH(z)}{H_0}}\right\}}
	-1,
	\end{align}
	 where the ratio $H^2(z)/H_0^2$ is given by (\ref{E2model2}).

We mention that while analyzing the model for the data set of $\Delta
\alpha/ \alpha$ of Table \ref{tab_data}, we have marginalized over $\Omega_{m0}$,
and thus the statistical information focuses only on the parameter $b$. For the fittings
$\Delta \alpha/ \alpha + SNIa$ and $\Delta \alpha/ \alpha + SNIa + BAO$ we have taken
$\Omega_m$ as a free parameter, and we note that $\Omega_{m0} = 0.277 \pm 0.019$
(for $\Delta \alpha/ \alpha$ $+$ SNIa) and $\Omega_{m0} = 0.283 \pm 0.016$
(for $\Delta \alpha/ \alpha$ $+$ SNIa $+$ BAO) at 1$\sigma$ confidence level.
\begin{figure}[ht]
 	\includegraphics[width=3.0in]{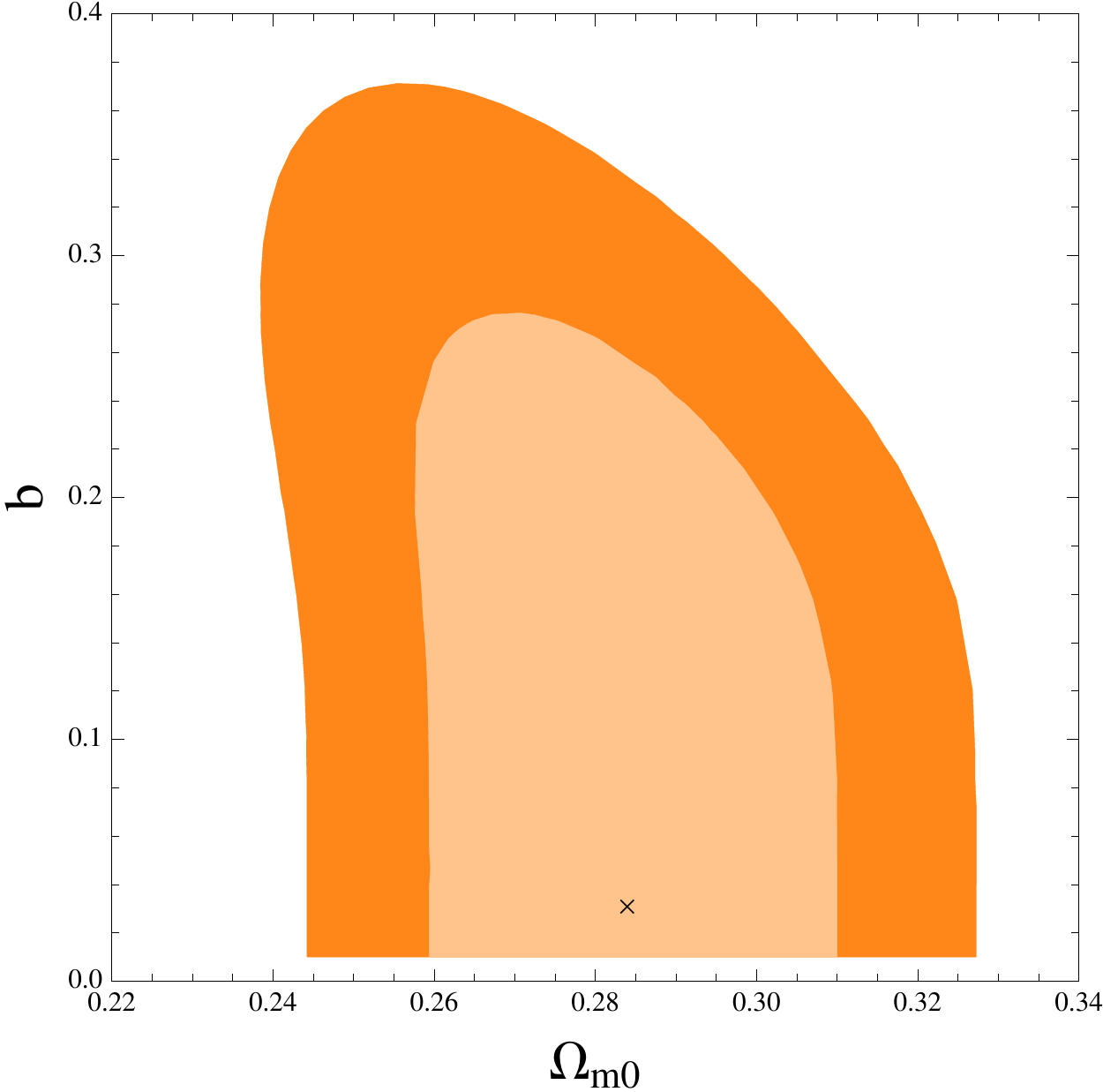}
 	\caption{\label{f2_Delta_alpha}  \textit{
 	1$\sigma$ and 2$\sigma$ confidence
regions for the $f_2$CDM  square-root-exponential  model of (\ref{model2}),  obtained
from the joint analysis   $\Delta \alpha/ \alpha$ $+$ SNIa $+$ BAO. The cross marks the
best-fit value.  }}
\end{figure}
Finally, in Fig. \ref{f2_Delta_alpha} we present the
68.27$\%$ and 95.45$\%$ confidence regions in the plane $\Omega_{m0} - b$,
considering the observational data $\Delta \alpha/ \alpha$  $+$ SNIa $+$ BAO (we have
taken $b \gtrsim 0.001$ in order to avoid divergences in the function $H(z)$ at high
redshifts). Note that
these results are in qualitative agreement with those of different observational fittings
\cite{Wu:2010mn,Nesseris:2013jea}, and show that   $\Lambda$CDM cosmology (which is
obtained for $b \rightarrow 0^{+}$) is inside the obtained region. 
Furthermore, and similarly to the $f_1$CDM model, from Table \ref{f2_1} we deduce 
that although the data from $ \Delta \alpha/ \alpha$ alone show a slightly deviating 
nature (reduced $\chi^2= 
1.1$) from $\Lambda$CDM scenario, but for $ \Delta \alpha/ \alpha$ $+$ SNIa and 
$\Delta \alpha/ \alpha$ $+$ SNIa $+$ BAO data it is implied that the model is very close 
to $\Lambda$CDM cosmology. This is also a self-consistent 
verification for the validity of our assumption that  $f(T)=T+const.+\text{corrections}$, 
which allowed us to work in the Einstein frame. 
\begin{table}[!h]
      \begin{center}
          \begin{tabular}{cccc}
          \hline
          \hline
               &$ Data $&$ b $& $\chi^2_{min}/d.o.f$\\
          \hline
          \hline
               &$ \Delta \alpha/ \alpha $                 &  $0.94  \pm 1.98$ & $1.1$ \\
               &$ \Delta \alpha/ \alpha$ $+$ SNIa         & $0.038 \pm 0.161$ & $0.97$   
\\
               &$ \Delta \alpha/ \alpha$ $+$ SNIa $+$ BAO & $0.031 \pm 0.246$ & $0.97$   
\\
          \hline
          \hline
          \end{tabular}
      \end{center}
      \caption{\textit{
 Summary of the best fit values of the parameter $b\equiv1/p$  of the
$f_2$CDM  square-root-exponential model of (\ref{model2}), for three
different observational data  sets with reduced $\chi^2$: $\chi^2_{min}/d.o.f$ ($d.o.f$
stands for ``degrees of freedom''). } }
      \label{f2_1}
\end{table}

\begin{figure}[ht]
 	\includegraphics[width=3.0in]{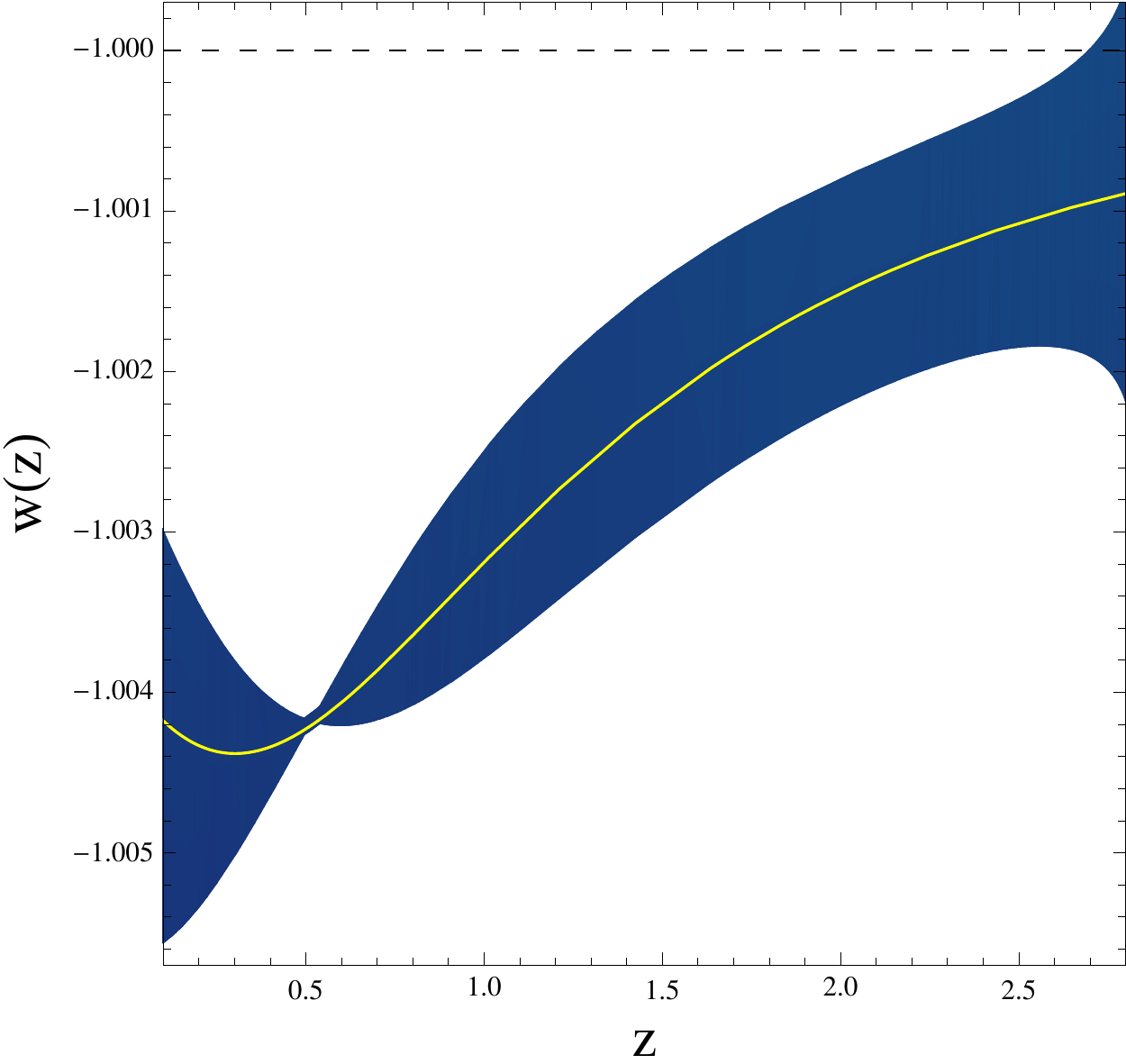}
 	\caption{\label{f2-totalEoS}  \textit{The evolution of the  
equation-of-state parameter given in (\ref{total-eos}),  for the $f_2$CDM 
 square-root-exponential model of (\ref{model2}), applying a reconstruction 
at 1$\sigma$ confidence level via error propagation using the joint analysis $\Delta 
\alpha/ \alpha$  $+$ SNIa $+$ BAO.
}}
\end{figure}

Lastly, in order to examine the late-time asymptotic behavior of  $f_2$CDM model, 
in Fig. \ref{f2-totalEoS} we depict the evolution of the  
equation-of-state parameter given in (\ref{total-eos}), applying a reconstruction 
at 1$\sigma$ confidence level via error propagation using the joint analysis $\Delta 
\alpha/ \alpha$  $+$ SNIa $+$ BAO, where one can see that $w$ at late times acquires 
values very  close to `$-1$', as expected. Similarly to the previous model, for a more 
detailed investigation of the  late-time asymptotics one must apply 
the method of dynamical system analysis \cite{Wu:2010xk}, where it can 
thoroughly be shown that the universe will end in a de Sitter phase.

\section{Observational constraints from the Newton's constant  variation}
\label{G}

\begin{figure*}[ht]
\includegraphics[width=6.9in, height=3.5in]{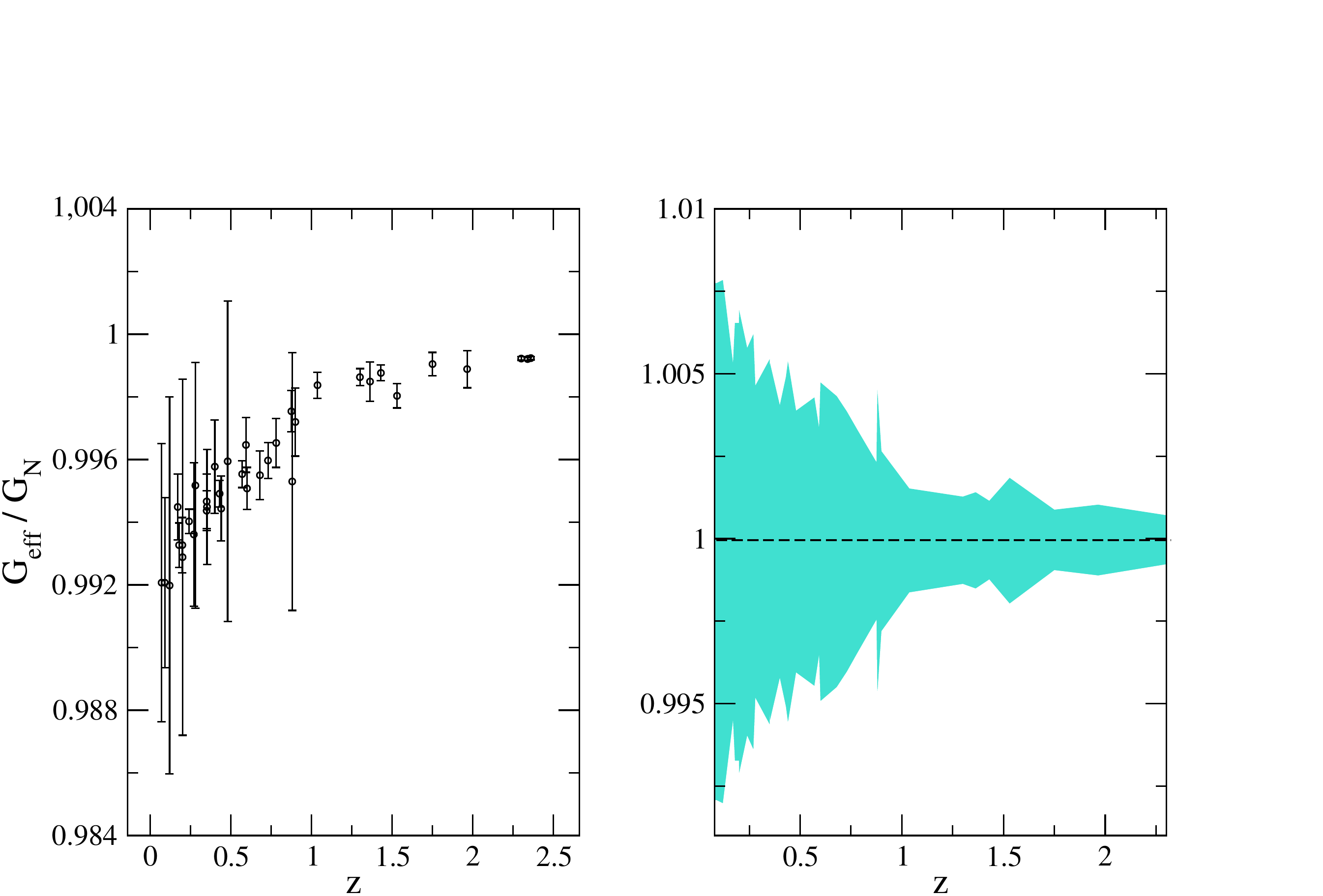}
\caption{\label{fT1_Geff}  \textit{Results for the $f_1$CDM  power-law model  of 
(\ref{model1}). Left graph: Estimation of $G_{eff}/G_N$ as a function of the redshift, 
from $37$ Hubble data points. Right graph: Reconstruction of $G_{eff}/G_N$ as a function 
of the redshift, 
from the observational Hubble parameter data, for $b$ $\in[-0.01,0.01]$.}}
\end{figure*}
In this section we will  directly use the observational constraints imposed on 
$f(T)$ models in order 
to examine the variation of the gravitational constant $G_N$. 
Let us first quantify
the $G_N$-variation in the framework of $f(T)$ cosmology. As is well known, a varying
effective gravitational constant is one of the common features in many modified
gravity theories \cite{Capozziello:2011et}. In case of $f(T)$ gravity, the effective
Newton's constant $G_{eff}$ can be straightforwardly extracted as
\cite{Zheng:2010am,Nesseris:2013jea}
\begin{equation}
\label{GNeff}
G_{eff} = \frac{G_N}{f_T}.
\end{equation}
Hence, for the   $f_1$CDM  power-law model of (\ref{model1}), and using (\ref{fcdm1}), we
obtain
 \begin{equation}
\label{GNeffmodel1}
G_{eff}(z) = \frac{G_N}{1- b\,\left(\frac{1-\Omega_{m0}}{2b-1} \right)\,
\left[\frac{H^2(z)}{H_0^2}\right]^{(b-1)}},
\end{equation}
where the ratio $H^2(z)/H_0^2$ is given by (\ref{E2model1}) (clearly, for $b= 0$ 
we have that   $G_{eff}(z)=G_N=const.$).
Similarly, for the $f_2$CDM  square-root-exponential  model of (\ref{model2}), and
using (\ref{fcdm2}),  we acquire
\begin{equation}
\label{GNeffmodel2}
G_{eff}(z) = \frac{G_N}{ 1+ \frac{p}{2} \left[ \frac{1-\Omega_{m0}}{1-(1+p)\,
e^{-p}}\right]\,
\left[ \frac{
H_0}{H(z)} \right] e^{-\frac{pH(z)}{H_0}}},
\end{equation}
 where the ratio $H^2(z)/H_0^2$ is given by (\ref{E2model2}) (clearly, for 
$b=1/p\rightarrow 0^+$ 
we have that   $G_{eff}(z)=G_N=const.$).

Let us now use the above expressions for $G_{eff}(z)$ and confront them with the
observational bounds of the Newton's constant variation. We use
observational Hubble parameter data in order to  investigate
the temporal evolution of the function $G_{eff}(z)$, since such a compilation is
usually used to constrain cosmological parameters,
due to the fact that  it is obtained from model-independent direct observations.
We adopt $37$ observational Hubble parameter data in the redshift range $0< z \leq
2.36$, compiled in \cite{Meng:2015loa}, out of which 27 data points are deduced from the
differential age method, whereas 10 correspond to measures obtained from the radial
baryonic acoustic oscillation method.

 We apply the following methodology:
Firstly, we estimate the error in the measurements associated with the function
$G_{eff}/G_N$, for 
both models of (\ref{GNeffmodel1}) and (\ref{GNeffmodel2}), via the standard method 
of error propagation theory, namely
\begin{eqnarray}
\label{GNeffmodel2bb}
&&\!\!\!\!\!\!\!\!\!\!\!\!\!\!\!
\sigma^2_{G_{eff}/G_N} = \Big|\frac{\partial G_{eff}/G_N}{\partial H}\Big|^2 \sigma^2_H 
+ \Big|\frac{\partial G_{eff}/G_N}{\partial b}\Big|^2 \sigma^2_b 
\nonumber\\
&&\ \ \ \ \ \ \ \ \ \ \ \ \ \ \ \ \ \ \ \ \ \ \ \ \ \ \ \  \ \ \ \ 
+
\Big|\frac{\partial G_{eff}/G_N}{\partial \Omega_m}\Big|^2 \sigma^2_{\Omega_m},
\end{eqnarray}
 and we  fix the free parameters of the two models within the values obtained in 
the joint analysis of \cite{Nunes:2016qyp} and of Section \ref{alpha} of the 
current work. Then, the measurements of $G_{eff}/G_N$ are 
calculated directly for each redshift defined in the adopted compilation.

\begin{figure*}[!]
\includegraphics[width=6.9in, height=3.5in]{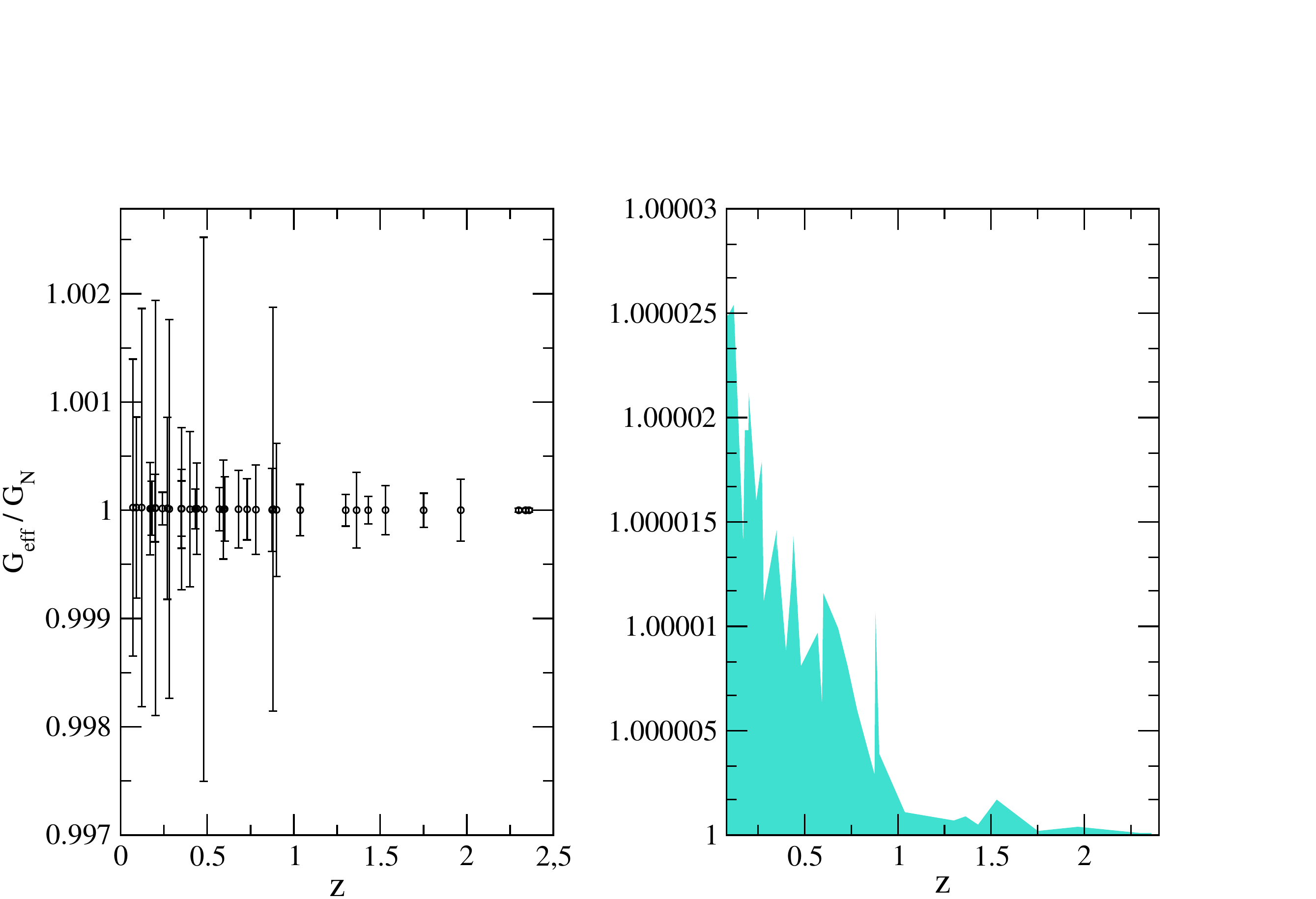}
\caption{\label{fT2_Geff}  \textit{
Results for the $f_2$CDM  square-root-exponential model  of 
(\ref{model2}). Left graph: Estimation of $G_{eff}/G_N$ as a function of the redshift, 
from $37$ Hubble data points. Right graph: Reconstruction of $G_{eff}/G_N$ as a function 
of the redshift, 
from the observational Hubble parameter data, for $b\equiv 1/p$ $\in[0,0.5]$. }}
\end{figure*}

In the left graph of Fig.  \ref{fT1_Geff} we depict  the  1$\sigma$ confidence-level
estimation of the function $G_{eff}(z)/G_N$ from $37$ Hubble data points, in the case of
the  $f_1$CDM  power-law model  of (\ref{model1}). Additionally, in the right graph of 
Fig. \ref{fT1_Geff} we present the corresponding  1$\sigma$ confidence-level  
reconstruction of $G_{eff}(z)/G_N$ for $b$ $\in [-0.01,0.01]$ from the observational 
Hubble parameter data. When we perform the analysis within the known range of the 
parameter $b$ for this model (from \cite{Nunes:2016qyp} as well as from Fig. 
\ref{f1_Delta_alpha_2} above), we find that $G_{eff}/G_N \approx 1$. Nevertheless, a 
minor 
deviation is observed for the fixed value of $b = 0.01$ (see the left graph of
Fig. \ref{fT1_Geff}). For instance, note that $G_{eff}(z=0.07)/G_N = 0.992 \pm 0.004$ 
and $G_{eff}(z=2.36)/G_N =0.99923 \pm 0.00005$, for the first and the last data points of 
the redshift interval [$0$, $2.36$], respectively.

In  Fig. \ref{fT2_Geff}  we present the corresponding graphs  for the  $f_2$CDM  
square-root-exponential  model  of (\ref{model2}). When we perform the analysis within 
the 
known range of the parameter $b\equiv1/p$ for this model (from \cite{Nunes:2016qyp} as 
well as from Fig. \ref{f2_Delta_alpha} above), we find that $G_{eff}/G_N \approx 1$, 
similarly to the case of $f_1$CDM  model.

In summary, from the analysis of this section, we verify the results of the previous 
section, namely that the parameter $b$ that quantifies the deviation  of both  $f_1$CDM 
and $f_2$CDM models from $\Lambda$CDM cosmology, is very close to zero. These results are 
in qualitative agreement with previous observational constraints on $f(T)$ gravity, 
according to which only small deviations are allowed, with  $\Lambda$CDM paradigm being 
inside the allowed region \cite{Iorio:2012cm,Wu:2010mn,Nesseris:2013jea,Nunes:2016qyp}.

\section{Conclusions}
\label{Conclusions}

In the present work we have used observations related to the variation of fundamental 
constants, in order to impose constraints on the viable and most used $f(T)$ gravity 
models.  In particular, since $f(T)$ gravity predicts a variation of the fine-structure 
constant, we used  the recent observational bounds of this variation, from   direct 
measurements obtained by 
different spectrographic methods, along with standard probes 
such as  Supernovae type Ia and baryonic acoustic oscillations, in order to constrain the 
involved model parameters of two viable and well-used  $f(T)$  models.

For both the $f_1$CDM  power-law model, as well as for  the $f_2$CDM  
square-root-exponential  model, we found that the parameter that quantifies the 
deviation from  $\Lambda$CDM cosmology can be slightly different than its $\Lambda$CDM 
value, nevertheless the best-fit value is very close to the $\Lambda$CDM one. 
Additionally,  since $f(T)$ gravity predicts a varying  effective gravitational constant, 
we quantified its temporal evolution with the use of the previously constrained   model 
parameters. For both  $f_1$CDM  and $f_2$CDM  models, we found that the deviation   from 
$\Lambda$CDM cosmology, is very close to zero. 

These 
results are in qualitative agreement with previous observational constraints on $f(T)$ 
gravity \cite{Wu:2010mn,Nesseris:2013jea,Nunes:2016qyp}, however they have been obtained 
through completely independent analysis. 
In summary, $f(T)$ gravity is consistent with 
observations, and thus it can serve as a candidate for modified gravity, although, as 
every modified gravity, it may  have only small deviation from $\Lambda$CDM cosmology, a 
feature that must be taken into account in any $f(T)$ model-building.\\

\section*{Acknowledgments}
The authors would like to thank P.~Brax for useful discussions. Additionally, 
they thank an anonymous referee for clarifying comments.
The work of SP is supported by the National Post-Doctoral Fellowship (File No: 
PDF/2015/000640)  
under the Science and Engineering Research Board (SERB), Govt. of India. This article is 
based upon work from COST Action ``Cosmology and Astrophysics Network for Theoretical 
Advances and 
Training Actions'', supported by COST (European Cooperation in Science and Technology).


\begin{thebibliography}{}


  \bibitem{Capozziello:2011et}
 S.~Capozziello and M.~De Laurentis,
 Phys.\ Rept.\ {\bf 509}, 167 (2011).

\bibitem{Copeland:2006wr}
  E.~J.~Copeland, M.~Sami and S.~Tsujikawa,
  Int.\ J.\ Mod.\ Phys.\ D {\bf 15}, 1753 (2006);
  Y.~F.~Cai, E.~N.~Saridakis, M.~R.~Setare and J.~Q.~Xia,
  Phys.\ Rept.\  {\bf 493}, 1 (2010).


\bibitem{Stelle:1976gc}
  K.~S.~Stelle,
  Phys.\ Rev.\ D {\bf 16}, 953 (1977).



 \bibitem{Nojiri:2010wj}
S.~Nojiri and S.~D.~Odintsov,
Phys.\ Rept.\ {\bf 505}, 59 (2011);



\bibitem{Nojiri:2005jg}
  S.~Nojiri and S.~D.~Odintsov,
  Phys.\ Lett.\ B {\bf 631}, 1 (2005);
  A.~De Felice and S.~Tsujikawa,
  Phys.\ Lett.\ B {\bf 675}, 1 (2009).



\bibitem{Naruko:2015zze}
  A.~Naruko, D.~Yoshida and S.~Mukohyama,
  Class.\ Quant.\ Grav.\  {\bf 33}, no. 9, 09LT01 (2016);
  E.~N.~Saridakis and M.~Tsoukalas,
  Phys.\ Rev.\ D {\bf 93}, no. 12, 124032 (2016).

\bibitem{deRham:2014zqa}
  C.~de Rham,
  Living Rev.\ Rel.\  {\bf 17}, 7 (2014).




\bibitem{ein28}
A. Einstein 1928, Sitz. Preuss. Akad. Wiss. p. 217; ibid p. 224.



 \bibitem{Hayashi79}
  K. Hayashi and T. Shirafuji,
  Phys. Rev. D \textbf{19}, 3524 (1979);
  Addendum-ibid. \textbf{24}, 3312 (1982).


 \bibitem{Pereira.book}
 R. Aldrovandi, J.G. Pereira,
{\it{Teleparallel Gravity: An Introduction}},
Springer, Dordrecht, 2013.





\bibitem{Bengochea:2008gz}
  G.~R.~Bengochea and R.~Ferraro,
  Phys.\ Rev.\ D {\bf 79}, 124019 (2009).


\bibitem{Linder:2010py}
  E.~V.~Linder,
  Phys.\ Rev.\ D \textbf{81}, 127301 (2010).





\bibitem{Dent:2011zz}
  S.~H.~Chen, J.~B.~Dent, S.~Dutta and E.~N.~Saridakis,
  Phys.\ Rev.\ D {\bf 83}, 023508 (2011);
  R.~J.~Yang,
  Eur.\ Phys.\ J.\ C {\bf 71}, 1797 (2011);
  J.~B.~Dent, S.~Dutta and E.~N.~Saridakis,
  JCAP {\bf 1101}, 009 (2011);
  M.~Li, R.~X.~Miao and Y.~G.~Miao,
  JHEP {\bf 1107}, 108 (2011);
    Y.~F.~Cai, S.~H.~Chen, J.~B.~Dent, S.~Dutta and E.~N.~Saridakis,
  Class.\ Quant.\ Grav.\  {\bf 28}, 215011 (2011);
 M.~H. Daouda, M.~E.~Rodrigues and M.~J.~S.~Houndjo,
  Eur.\ Phys.\ J.\ C {\bf 72}, 1890 (2012).



\bibitem{Geng:2011aj}
  Y.~P.~Wu and C.~Q.~Geng,
     Phys.\ Rev.\ D {\bf 86}, 104058 (2012);
  K.~Bamba, R.~Myrzakulov, S.~Nojiri and S.~D.~Odintsov,
  Phys.\ Rev.\ D {\bf 85}, 104036 (2012);
  K.~Karami and A.~Abdolmaleki,
 JCAP 1204 (2012) 007;
  V.~F.~Cardone, N.~Radicella and S.~Camera,
  Phys.\ Rev.\ D {\bf 85}, 124007 (2012);
  G.~Otalora,
  JCAP {\bf 1307}, 044 (2013);
  J.~Amoros, J.~de Haro and S.~D.~Odintsov,
     Phys.\ Rev.\ D {\bf 87}, 104037 (2013).


\bibitem{Bamba:2013jqa}
  K.~Bamba, S.~Capozziello, M.~De Laurentis, S.~'i.~Nojiri and D.~S\'aez-G\'omez,
  Phys.\ Lett.\ B {\bf 727}, 194 (2013);
  S.~Bahamonde, C.~G.~B\"{o}hmer and M.~Wright,
  Phys.\ Rev.\ D {\bf 92}, 104042 (2015);
J.~de Haro and J.~Amoros,
  Phys.\ Rev.\ Lett.\  {\bf 110}, no. 7, 071104 (2013);
    M.~Kr\v{s}\v{s}\'ak and E.~N.~Saridakis,
  Class.\ Quant.\ Grav.\  {\bf 33}, no. 11, 115009 (2016);
A.~Paliathanasis, J.~D.~Barrow and P.~G.~L.~Leach,
  Phys.\ Rev.\ D {\bf 94}, no. 2, 023525 (2016).
  


 \bibitem{Cai:2015emx}
  Y.~F.~Cai, S.~Capozziello, M.~De Laurentis and E.~N.~Saridakis,
   Rept.\ Prog.\ Phys.\  {\bf 79}, no. 10, 106901 (2016).




\bibitem{Kofinas:2014owa}
  G.~Kofinas and E.~N.~Saridakis,
  Phys.\ Rev.\ D {\bf 90}, 084044 (2014);
  G.~Kofinas and E.~N.~Saridakis,
  Phys.\ Rev.\ D {\bf 90}, 084045 (2014).





\bibitem{Otalora:2016dxe}
  G.~Otalora and E.~N.~Saridakis,
   Phys.\ Rev.\ D {\bf 94}, no. 8, 084021 (2016).




 \bibitem{Iorio:2012cm}
   L.~Iorio and E.~N.~Saridakis,
  Mon. Not. Roy. Astron. Soc. 427 (2012) 1555;
  M.~L.~Ruggiero and N.~Radicella,
  Phys.\ Rev.\ D {\bf 91}, 104014 (2015);
L.~Iorio, N.~Radicella and M.~L.~Ruggiero,
   JCAP {\bf 1508} (2015) no.08,  021;
   G.~Farrugia, J.~L.~Said and M.~L.~Ruggiero,
   Phys.\ Rev.\ D {\bf 93}, no. 10, 104034 (2016).





\bibitem{Wu:2010mn}
   P.~Wu, H.~W.~Yu,
   Phys.\ Lett.\ \textbf{B693}, 415 (2010);
   S.~Capozziello, O.~Luongo and E.~N.~Saridakis,
   Phys.\ Rev.\ D {\bf 91}, no. 12, 124037 (2015).
    %



 \bibitem{Nesseris:2013jea}
   S.~Nesseris, S.~Basilakos, E.~N.~Saridakis and L.~Perivolaropoulos,
   Phys.\ Rev.\ D {\bf 88}, 103010 (2013).


\bibitem{Nunes:2016qyp}
  R.~C.~Nunes, S.~Pan and E.~N.~Saridakis,
  JCAP {\bf 1608}, no. 08, 011 (2016).

















\bibitem{dirac} P. A. M. Dirac, Nature \textbf{139} (1937) 323; Proc. Roy. Soc. London A
\textbf{
165} (1938) 198.

\bibitem{milne-jordan}  E. A. Milne, Relativity, Gravitation and World Structure
(Clarendon press,
Oxford, 1935);
Proc. Roy. Soc. \textbf{A3} (1937) 242; P. Jordan, Naturwiss. \textbf{25} (1937) 513; Z.
Physik 113
(1939)
660.


\bibitem{brans-dicke} P. Jordan, Nature \textbf{164}, 637 (1949); C. Brans, R.H. Dicke,
Phys. Rev.
D \textbf{124}, 925 (1961).

\bibitem{Gamow}
G. Gamow, Phys. Rev. Lett. \textbf{19}, 759 (1967).






\bibitem{Dvali:2001dd}
  G.~R.~Dvali and M.~Zaldarriaga,
  Phys.\ Rev.\ Lett.\  {\bf 88}, 091303 (2002);
  T.~Chiba and K.~Kohri,
  Prog.\ Theor.\ Phys.\  {\bf 107}, 631 (2002);
  L.~Anchordoqui and H.~Goldberg,
  Phys.\ Rev.\ D {\bf 68}, 083513 (2003);
  C.~Wetterich,
  Phys.\ Lett.\ B {\bf 561}, 10 (2003);
  E.~J.~Copeland, N.~J.~Nunes and M.~Pospelov,
  Phys.\ Rev.\ D {\bf 69}, 023501 (2004).

\bibitem{Bento:2004jg}
  M.~d.~C.~Bento, O.~Bertolami and N.~M.~C.~Santos,
  Phys.\ Rev.\ D {\bf 70}, 107304 (2004);
  V.~Marra and F.~Rosati,
  JCAP {\bf 0505}, 011 (2005);
  P.~P.~Avelino,
  Phys.\ Rev.\ D {\bf 78}, 043516 (2008);
  H.~B.~Sandvik, J.~D.~Barrow and J.~Magueijo,
  Phys.\ Rev.\ Lett.\  {\bf 88}, 031302 (2002);
  D.~F.~Mota and J.~D.~Barrow,
  Mon.\ Not.\ Roy.\ Astron.\ Soc.\  {\bf 349}, 291 (2004);
  J.~D.~Barrow, D.~Kimberly and J.~Magueijo,
  Class.\ Quant.\ Grav.\  {\bf 21}, 4289 (2004);
  H.~Wei,
  Phys.\ Lett.\ B {\bf 682}, 98 (2009);
  H.~Wei, X.~P.~Ma and H.~Y.~Qi,
  Phys.\ Lett.\ B {\bf 703}, 74 (2011).







\bibitem{Magueijo:2003gj}
  J.~Magueijo,
  Rept.\ Prog.\ Phys.\  {\bf 66}, 2025 (2003);
  J.~P.~Uzan,
  Living Rev.\ Rel.\  {\bf 14}, 2 (2011).




\bibitem{Avelino:2006gc}
  P.~P.~Avelino, C.~J.~A.~P.~Martins, N.~J.~Nunes and K.~A.~Olive,
  Phys.\ Rev.\ D {\bf 74}, 083508 (2006);
  E.~Garcia-Berro, J.~Isern and Y.~A.~Kubyshin,
  Astron.\ Astrophys.\ Rev.\  {\bf 14}, 113 (2007);
  J.~D.~Barrow,
  Annalen Phys.\  {\bf 19}, 202 (2010);
  T.~Chiba,
  Prog.\ Theor.\ Phys.\  {\bf 126}, 993 (2011);
  J.~K.~Webb, J.~A.~King, M.~T.~Murphy, V.~V.~Flambaum, R.~F.~Carswell and
M.~B.~Bainbridge,
  Phys.\ Rev.\ Lett.\  {\bf 107}, 191101 (2011);
   J.~A.~King, J.~K.~Webb, M.~T.~Murphy, V.~V.~Flambaum, R.~F.~Carswell,
M.~B.~Bainbridge, M.~R. ~Wilczynska and F.~E.~Koch,
   Mon.\ Not.\ Roy.\ Astron.\ Soc.\  {\bf 422}, 3370 (2012).






\bibitem{Sola:2015xga}
  J.~Sol\`{a},
  Mod.\ Phys.\ Lett.\ A {\bf 30}, no. 22, 1502004 (2015);
  A.~M.~M.~Pinho and C.~J.~A.~P.~Martins,
  Phys.\ Lett.\ B {\bf 756}, 121 (2016);
  H.~Fritzsch, R.~C.~Nunes and J.~Sola,
  arXiv:1605.06104 [hep-ph].















  \bibitem{Stadnik:2014tta}
   Y.~V.~Stadnik and V.~V.~Flambaum,
   Phys.\ Rev.\ Lett.\  {\bf 114}, 161301 (2015);
  W.~W.~Zhu {\it et al.},
  Astrophys.\ J.\  {\bf 809}, no. 1, 41 (2015).



\bibitem{Dvali:2000hr}
  G.~R.~Dvali, G.~Gabadadze and M.~Porrati,
  Phys.\ Lett.\ B {\bf 485}, 208 (2000).




 \bibitem{Olive:2001vz}
  K.~A.~Olive and M.~Pospelov,
  Phys.\ Rev.\ D {\bf 65}, 085044 (2002).



\bibitem{Brax:2012gr} 
  P.~Brax, A.~C.~Davis, B.~Li and H.~A.~Winther,
  Phys.\ Rev.\ D {\bf 86}, 044015 (2012);
  P.~Brax, C.~Burrage, A.~C.~Davis, D.~Seery and A.~Weltman,
  Phys.\ Lett.\ B {\bf 699}, 5 (2011).



\bibitem{Yang:2010ji}
  R.~J.~Yang,
  Europhys.\ Lett.\  {\bf 93}, 60001 (2011) 



  
  
  
  
  
  
  
  
  
  
  
\bibitem{Songaila:2014fza}
  A.~Songaila and L.~L.~Cowie,
  Astrophys.\ J.\  {\bf 793}, 103 (2014).

  
  
  
\bibitem{Evans:2014yva}
  T.~M.~Evans {\it et al.},
  Mon.\ Not.\ Roy.\ Astron.\ Soc.\  {\bf 445}, no. 1, 128 (2014).

  
  

 
   
  
\bibitem{Molaro:2007kp}
  P.~Molaro, D.~Reimers, I.~I.~Agafonova and S.~A.~Levshakov,
  Eur.\ Phys.\ J.\ ST {\bf 163}, 173 (2008).

\bibitem{Chand:2006va}
  H.~Chand, R.~Srianand, P.~Petitjean, B.~Aracil, R.~Quast and D.~Reimers,
  Astron.\ Astrophys.\  {\bf 451}, 45 (2006).

\bibitem{Molaro:2013saa}
  P.~Molaro {\it et al.},
  Astron.\ Astrophys.\  {\bf 555}, A68 (2013).
  
  

\bibitem{Agafonova:2011sp}
  I.~I.~Agafonova, P.~Molaro, S.~A.~Levshakov and J.~L.~Hou,
  Astron.\ Astrophys.\  {\bf 529}, A28 (2011).


  
  
  
\bibitem{Suzuki:2011hu}
  N.~Suzuki {\it et al.},
  Astrophys.\ J.\  {\bf 746}, 85 (2012).

\bibitem{Blake:2011en}
  C.~Blake {\it et al.},
  Mon.\ Not.\ Roy.\ Astron.\ Soc.\  {\bf 418}, 1707 (2011).

  
  
  
  
  
  
  
  
  
  
  
  
\bibitem{Shi:2012ma}
  K.~Shi, Y.~Huang and T.~Lu,
  Mon.\ Not.\ Roy.\ Astron.\ Soc.\  {\bf 426}, 2452 (2012).

     
     
\bibitem{Wu:2010xk} 
  P.~Wu and H.~W.~Yu,
  Phys.\ Lett.\ B {\bf 692}, 176 (2010);
  C.~Xu, E.~N.~Saridakis and G.~Leon,
  JCAP {\bf 1207}, 005 (2012);
  G.~Kofinas, G.~Leon and E.~N.~Saridakis,
  Class.\ Quant.\ Grav.\  {\bf 31}, 175011 (2014).
     
     
     
     
     
     
     
     
     
     
     
     
     
     
\bibitem{Zheng:2010am}
   R.~Zheng and Q.~G.~Huang,
   JCAP {\bf 1103}, 002 (2011).

   
   

    
  
  
\bibitem{Meng:2015loa}
  X.~L.~Meng, X.~Wang, S.~Y.~Li and T.~J.~Zhang,
  arXiv:1507.02517 [astro-ph.CO].


  


  
 




  
  

\end{thebibliography}
\end{document}